\documentclass[prl,twocolumn,showpacs]{revtex4}

\usepackage{graphicx}
\usepackage{amsmath}

\newcommand{\abs}[1]{\left| #1 \right|}
\newcommand{\eq}[1]{Eq.~(\ref{#1})}
\newcommand{\twoEq}[2]{Eqs.~(\ref{#1}) and (\ref{#2})}

\newcommand{\fig}[1]{Fig.~\ref{#1}}

\newcommand{\mypartial}[2]{\frac{\partial #1}{\partial #2}}

\newcommand{\trace}[2]{\text{tr}_{#1}\left\{{#2}\right\}}

\def \be{\begin{equation}}
\def \ee{\end{equation}}
\def \bmlett{\begin{mathletters}}
\def \emlett{\end{mathletters}}

\def \e{\epsilon}

\def \HH{{\mathcal H}}

\def \RR{{\mathcal R}}

\def \TT{{\mathcal T}}

\def \Hint {\HH_\text{int}}

\def \adag {a^\dagger}

\def \Teff {T_{\text{eff}}}

\def \ed {\e_d}

\def \Gtot {\Gamma}
\def \re {\vec{r}_e}
\def \xe {x_e}

\def \aop {\hat a}
\def \adag {\hat a^\dagger}
\def \Te {T_\text{el}}

\begin{document}

\title{Scattering approach to
backaction in coherent nanoelectromechanical systems}
\author{Steven D. Bennett}
\author{Jesse Maassen}
\author{Aashish A. Clerk}
\date{\today}
\affiliation{Department of Physics, McGill University,
Montreal, Quebec, Canada, H3A 2T8}
 
\pacs{85.85.+j, 73.23.-b, 72.70.+m}

\begin{abstract}
We present theoretical results for 
the backaction force noise and damping
of a mechanical oscillator whose position is measured by
a mesoscopic conductor.
Our scattering approach is applicable to 
a wide class of systems;
in particular,
it may be used to describe
point contact position detectors far from the 
weak tunneling limit.
We find 
that the backaction depends not
only on the mechanical modulation of 
transmission
probabilities
but also on the modulation of scattering phases,
even
in the absence of a
magnetic field.
We illustrate our general approach with several simple examples,
and use it to calculate the backaction for 
a movable, Au atomic point contact modeled by {\it ab initio}
density functional theory.
\end{abstract}

\maketitle

Quantum mechanics requires that any detector used
to 
measure an object's position unavoidably
exerts a backaction force,
imposing a fundamental limit on 
continuous position 
detection  \cite{Caves82,ClerkRMP}.  
Recent 
experiments with nanoelectromechanical 
systems (NEMS) have come 
remarkably close to realizing this limit by using 
quantum electronic conductors as 
position detectors of nanomechanical 
oscillators \cite{Knobel03,Naik06,Flowers-Jacobs07}.
In these systems, position detection is
achieved using
the influence of
the mechanics on the current 
through the conductor; thus, 
it is natural to associate
backaction
with the 
position sensitivity 
of the electron transmission probability.
This is indeed the picture that emerges
from theoretical 
studies in the limit of weak 
tunneling \cite{Mozyrsky02,ClerkGirvin04,Doiron08};
however, several 
recent experiments are far from this 
limit \cite{Flowers-Jacobs07,Kemiktarak07,Poggio08,Stettenheim10},
and it is not clear that the weak tunneling results
apply.

In this paper, we study the backaction
of a mesoscopic position detector 
using a general noninteracting
scattering approach that 
is not limited to the weak 
tunneling limit.
Scattering theory has been
used extensively to
study various aspects of mesocopic
conductors, 
and we adapt it here to study the 
backaction heating and damping
of a mechanical oscillator coupled to
a conductor.
Surprisingly, 
we find that backaction arises not only
from transmission probabilities,
but also from the position sensitivity of
scattering phases,
and present several simple but
illustrative 
examples where the phases 
play a pronounced
role.
We emphasize that these phases may
be important despite intact
time reversal symmetry,
which
we assume throughout,
unlike Aharanov-Bohm phases due
to a magnetic field \cite{Doiron08}.
Finally, we apply our general results
to calculate the backaction from
an atomic point contact (APC) 
between Au electrodes, using a
scattering matrix 
obtained from density functional theory (DFT).

Our approach 
significantly extends the seminal
work of Yurke and Kochanski, 
who first
considered force noise
in a tunnel junction
using a scattering approach \cite{Yurke90}.
Unlike their work, which
is limited to particular
scattering potentials,
we rely only on general
properties of the
scattering matrix.
As a result, we can describe a
wide class of systems
including arbitrary scattering
potentials,
various forms
of electromechanical coupling, 
and multichannel scatterers.
Moreover, we calculate
not only the backaction force noise but also
the backaction damping,
which
is important in experiments
(e.g.~it is the basis of backaction-cooling
\cite{Naik06}) and
up to now has not been dealt with
in the scattering approach.

{\it Scattering approach.}---We 
consider a two-terminal device
consisting of a coherent
scattering region coupled to left and right
leads, each of which supports $N$ 
transverse modes.
Electrons are scattered
by a potential $U(\re,x)$,
where $\re = (\xe, y_e, z_e)$ is the
electron position, and the potential
depends on the
position $x$ of a mechanical
oscillator.
Incoming and outgoing
waves are related
by the scattering matrix $s(x)$, 
which depends on $x$
through $U(\re,x)$.
We will show that a knowledge of
$s$ and $\partial s / \partial x$
is sufficient to calculate the backaction.

For the usual
experimental regime of
weak electromechanical coupling,
the change in the electronic potential
due to small changes in $x$
are generally linear and may be written
$\HH_\text{int} = - x \hat{F}$,
where the force on the
oscillator
is $\hat{F} = -  \int d\re \hat{\rho}(\re)
\partial U(\re,x) / \partial x$,
and $\hat\rho(\re)$ is the electron
density operator.
By relating small and
slow changes
in the potential, $U(\vec r_e,x)$,
to the parametric derivative
of the scattering matrix \cite{Buttiker93,Aleiner02,Mahaux68},
we can express $\hat F$ in
the scattering state basis as
\begin{equation}
\label{eq:F}
\hat{F} =  
		\sum_{\alpha\beta}  \int d\e \int d\e'
		\adag_{\alpha}(\e) W_{\alpha\beta}(\e,\e') \aop_{\beta}(\e'),
\end{equation}
where
$\aop_{\alpha}(\e)$
destroys a scattering state 
 of energy $\e$ 
incident in lead $\alpha$,
and
\begin{equation}
\label{eq:W}
	W(\e,\e)
	= 
	\frac{1}{2\pi i} \left[ s^\dagger(\e,x) \mypartial{s(\e,x)}{x} \right]_{x=0}.
\end{equation}
We require only the diagonal-in-energy part of $W$
since we focus on the zero frequency noise properties of $\hat F$,
sufficient for the experimentally relevant case when
the oscillator period is much longer
than timescales in the conductor.
Derivatives of the scattering matrix similar to
\eq{eq:W} are familiar from 
studies of charge noise \cite{Buttiker98}
and parametric pumping  \cite{Brouwer98}.
Here we use the parametric
derivative with respect to $x$ to calculate the backaction 
on the oscillator 
directly in terms of the
scattering matrix, without the need for a detailed knowledge
of $U(\re,x)$ and $\hat\rho(\re)$ in the scattering region.
In the following we work to lowest order in $\Hint$,
valid for weak coupling.

Fluctuations of the backaction
force 
cause momentum diffusion and heating 
of the oscillator.
Heating is determined by
the classical,  
frequency-symmetric
part of the backaction force noise,
$\bar S_F[\omega] = \left( S_F[\omega] + S_F[-\omega] \right)/2$,
where the quantum
noise spectral density
is $S_F[ \omega] = \int dt e^{i  \omega t} 
\langle \hat{F}(t) \hat{F}(0) \rangle$
and averages are taken with respect to the 
uncoupled conductor \cite{ClerkBennett05}.
These averages are easily taken using \eq{eq:F},
and the backaction heating is directly determined
by $W$.
The zero frequency force noise is 
($k_B = 1$, $\bar S_F \equiv \bar S_F[0]$)
\begin{equation}
\label{eq:SFgen}
	\bar S_{F} = 2\pi\hbar
	\sum_{\alpha\beta}
	\int d\e
	 \trace{}{ W_{\alpha\beta} W_{\beta\alpha} } f_\alpha \left( 1 - f_\beta \right),
\end{equation}
where the trace is over transverse modes,
assumed to be the same in both leads,
and the matrixes $W_{\alpha\beta}$ are the
$N \times N$
blocks of $W$ in \eq{eq:W}, which may be
$\e$-dependent.
The Fermi functions are
$f_\alpha = (1 + e^{(\epsilon-\mu_\alpha)/\Te})^{-1}$,
where $\mu_\alpha$ is the chemical potential
in lead $\alpha$ and $\Te$ is the electronic temperature . 

In addition to heating, the
oscillator also experiences backaction damping
as a result of energy exchange
with the conductor.
The damping rate is
given by the quantum,
asymmetric-in-frequency
part of the force noise,
$\gamma[\omega] = 
\left( S_F[\omega] - S_F[-\omega] \right) / 2M\hbar\omega$,
where $M$ is the oscillator mass.
Taking the $\omega \rightarrow 0$ limit, we find
\begin{equation}
\label{eq:dampingGen}
	\gamma = \frac{2\pi \hbar}{M}
	\sum_{\alpha\beta}
	\int d\e
	\trace{}{ W_{\alpha\beta} W_{\beta\alpha} } f_\alpha
	\left( - \mypartial{f_\beta}{\e} \right).
\end{equation}
By considering the ratio of 
$\bar S_F[\omega]$ to 
$\gamma[\omega]$,
one can associate a frequency-dependent 
effective temperature, $\Teff[\omega]$,
with the 
backaction;
this amounts to using the 
standard fluctuation-dissipation relation 
to {\it define} the effective temperature at 
each frequency from the system's force 
noise and damping 
\cite{ClerkRMP, Mozyrsky02, ClerkBennett05}.
$\Teff[\omega]$
characterizes the conductor
as an effective thermal environment.
In the $\omega \rightarrow 0$ limit,
the relation is simply
$\Teff \equiv \bar{S}_F / {2 M \gamma}$.
If backaction dominates
over intrinsic sources of dissipation,
$\Teff$ corresponds to the physical
temperature of the oscillator.

{\it Single channel.}---We first 
consider the case of single-channel leads.
For simplicity, we also focus
on the limit of small applied bias,
ignoring the possible 
energy-dependence of $s$.
We assume
time reversal
symmetry (i.e.~no magnetic field), 
but allow for broken left-right
inversion symmetry.
In this case the scattering
matrix may be parametrized as
\begin{equation}
\label{eq:s}
	s(\e,x) = e^{i\phi}
	\begin{pmatrix}
		\sqrt{\RR} e^{i \theta} & i \sqrt{\TT} \\
		i \sqrt{\TT} & \sqrt{\RR} e^{-i \theta }
	\end{pmatrix},
\end{equation}
where $\TT$ ($\RR = 1-\TT$) 
is the transmission (reflection) probability,
$\phi$ is the overall scattering phase,
and $\theta$ parametrizes
broken inversion symmetry,
i.e.~$\theta=0$ for
an inversion-symmetric conductor.
In general,
all of the scattering parameters
depend on $x$ through the potential,
$U(\re,x)$.

Inserting \eq{eq:s} into \eq{eq:SFgen}
we obtain the symmetrized force noise
for a single channel,
\begin{align}
\label{eq:SF}
	\bar{S}_{F} &= \frac{\hbar}{2\pi} 
	\frac{ (\partial \TT / \partial x)^2 }{4\RR\TT} eV
	\\	
	& \times \left[	
	 \left( 1 + \RR\TT \Delta_\theta \right)
	\coth{\left(\frac{eV}{2\Te}\right) } 
	+ \left( \Delta_\phi
	+ \RR^2 \Delta_\theta \right) \frac{2\Te}{eV} \right],
	\nonumber
\end{align}
where $V$ is the bias, and
the phase terms enter as
\begin{equation}
\label{eq:phaseCorrection}
	\Delta_\zeta = 4\RR\TT 
	\left( \frac{\partial \zeta / \partial x}{\partial \TT/\partial x} \right)^2,
\end{equation}
for $\zeta = \phi,\theta$.
In the limit $eV \gg \Te$,
the first term
(independent of $\Delta_{\phi}$ and $\Delta_\theta$)
in \eq{eq:SF} 
represents the expected, 
quantum-limited backaction of our position detector:  
it is simply the sensitivity of a position measurement
by monitoring the current, 
and reflects the fact that a stronger 
measurement leads to increased backaction.  
This term scales as the square of the 
measurement gain, $\chi_{IF} \propto \partial \TT / \partial x$, and inversely 
with the shot noise in the current, 
$\bar S_I = e^2 V \RR\TT / 2\pi\hbar$;
in the limit $\TT \ll 1$ it
reproduces the well-known result 
obtained from a tunnel Hamiltonian calculation 
\cite{Mozyrsky02,ClerkGirvin04}.
The second term in \eq{eq:SF} is 
independent of $\partial \TT / \partial x$
and is thus not directly related to a measurement
of the current.
Instead, it results from the
oscillator's modulation of the phase 
$\theta$.
This phase contribution 
to $\bar S_F$
is proportional to $\RR\TT$
and thus vanishes when $\TT \ll 1$.
The remaining two terms ($\propto \Te / eV$)
are also independent of $\partial \TT / \partial x$ and
represent additional thermal noise
at finite $\Te / eV$.

The damping for a single channel is 
\begin{equation}
\label{eq:damping}
	\gamma = \frac{\hbar}{2\pi M}
	\frac{ (\partial \TT / \partial x)^2 }{4\RR\TT} 
	\big( 1 + \Delta_\phi
	+ \RR \Delta_\theta
	 \big).
 \end{equation}
In the small bias limit, $\gamma$ is
strictly positive and independent of $\Te$.
Similar to $\bar S_F$,
the first term in \eq{eq:damping}
is the backaction
associated with a measurement of the current
and
reduces to the tunnel Hamiltonian
result
in the limit $\TT \ll 1$.
More interestingly, the second and third
terms correspond to
corrections due to scattering phases;
unlike $\bar{S}_F$, these phase
contributions to $\gamma$ are present even 
for a symmetric detector
and, as we will see,
do not necessarily vanish in
the weak tunneling limit.
The overall phase $\phi$
is directly connected
to the density of
states in the scattering region
via the 
Friedel sum rule \cite{Friedel52}.
An $x$-dependent $\phi$ implies 
that the mechanical oscillator can 
change the scattering-induced electronic density of states;
this means that the total electronic
free energy becomes $x$-dependent, resulting
in a force whose
quantum noise contributes to
damping.

Equations (\ref{eq:SF}) and (\ref{eq:damping})
show that the backaction properties of a general
conductor cannot simply be extrapolated from
the weak tunneling limit; scattering phases
play a role in both the heating and damping
of the oscillator.
Further, the phases can have a dramatic
influence on the effective backaction temperature 
$\Teff$ of the detector.
For a single channel,
using \twoEq{eq:SF}{eq:damping},
in the limit $\Te \ll eV$ we find
\begin{equation}
\label{eq:Teff}
	\Teff = \frac{eV}{2}
	\left(
	  \frac{ 1 + \RR\TT \Delta_\theta}
	  {1 + \Delta_\phi + \RR \Delta_\theta}
	\right).
\end{equation}
If 
the mechanical motion does not modify 
the scattering phases, 
then we simply obtain 
the tunnel Hamiltonian 
result \cite{Mozyrsky02,ClerkGirvin04},
$\Teff = eV/2$,
independent of $\TT$.
However, 
in the more general case including the
backaction from scattering phases, 
$\Teff$ is not
solely determined by the voltage.
The phase corrections
always decrease the effective temperature;
they arise from the diagonal elements
of $W$,
which correspond to transitions between
scattering states in
the same lead.
At $\Te = 0$ such transitions can only
occur if an
electron absorbs energy, because
the scattering states in each lead are filled
up to the Fermi level.
Thus, phase corrections 
lead to increased absorption of energy 
from the oscillator, lowering $\Teff$.
Including a non-zero lead temperature $\Te$, 
one finds that 
$\Teff$ can be lowered to a minimum value of $\Te$; 
as $\Te \ll eV$, 
this could still be quite useful.

{\it Square potential barrier.}---To demonstrate
that backaction from scattering phases
plays a role even in the simplest
scattering model, we calculate the backaction for
a one dimensional symmetric square barrier potential
whose width depends on the
oscillator position.
The force noise for this model was
first considered in Ref.~\onlinecite{Yurke90};
our general method further provides 
$\gamma$ and $\Teff$ and allows us
to identify the role of scattering phases. 
Incoming electrons of wavevector $k$
and energy $\e$ 
are scattered in one dimension
by a square potential barrier of height $U_0$
and width $w = L + x$.
The inverse decay length of the wavefunction under the
barrier is 
$\kappa = \sqrt{2m_e(U_0 - E)}/\hbar$.
It is straightforward to find $\TT$ and $\phi$
as functions of $U_0$ and $w$, 
and $\theta = 0$
due to inversion symmetry.
We obtain
\begin{equation}
	\Delta_\phi
	= \left( 1 + \frac{4k^2\kappa^2}{\left( k^2-\kappa^2 \right)^2 \TT}
	\right)^{-1},
\end{equation}
and the phase terms become important 
when $\TT \sim 1$.
For a high but narrow barrier ($U_0 \gg \e$, $\kappa L \ll 1$),
we find (via \eq{eq:Teff}) that $\Teff$ may be reduced
by up to a factor two 
compared to the tunnel Hamiltonian result of $eV/2$.
For a low barrier ($U_0 \ll \e$), we find
$\Teff \rightarrow \Te$.

\begin{figure}[tb]
\centering
	\hspace{5mm}
	\includegraphics[width=0.42\textwidth]{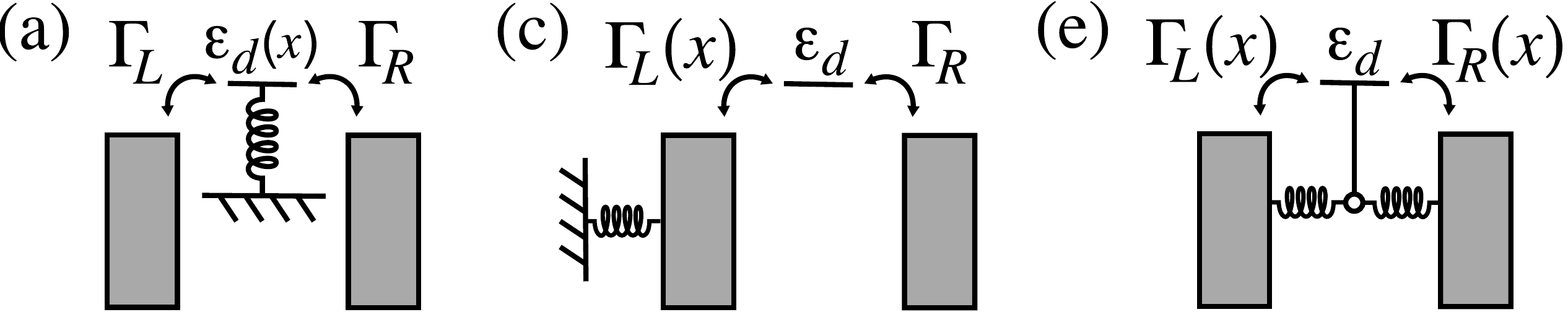}
	\\
	\includegraphics[width=0.45\textwidth]{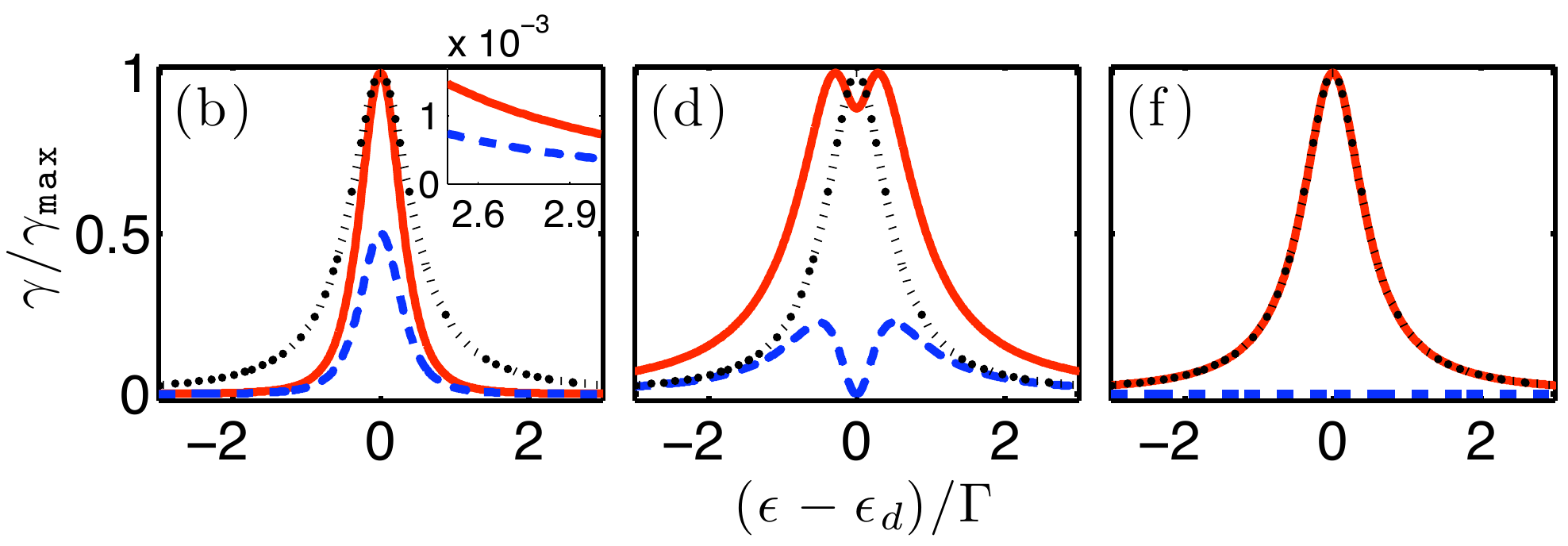}
\caption{
Schematic setups and backaction damping
for RLM with dispersive (a,b),
dissipative (c,d) and shuttle (e,f) 
mechanical coupling.
In all plots,
the full backaction damping (red solid)
and the damping without phase corrections 
(blue dashed) are shown,
with $\TT$ (black dotted) also shown 
for reference.  
We took
$\Gamma_L(0) = \Gamma_R(0)$.
}
\label{fig:rlm}
\end{figure}

{\it Resonant level model.}---We
now apply our general results to a prototypical
resonant level model
(RLM), 
where a single electronic level
of energy $\ed$ is
connected to the left (right) lead via tunneling
rate $\Gamma_L$ ($\Gamma_R$).
If $\ed$ depends on the 
position of a mechanical oscillator
(see \fig{fig:rlm}a),
one has the electromechanical
analog of a dispersively coupled optomechanical
system \cite{Marquardt07}, and
a simple model of quantum-dot-based 
NEMS studied in recent
experiments \cite{Steele09,Lassagne09}.
Beginning from the scattering
matrix for the RLM,
$s_{\alpha\beta}
	= \delta_{\alpha\beta} - 
	i \hbar \sqrt{ \Gamma_\alpha \Gamma_\beta} /
	\left(  \e - \ed + i \hbar \Gtot/2 \right)$,
where $\Gamma = \Gamma_L + \Gamma_R$,	
and assuming linear coupling,
we obtain 
$\gamma \propto (\Gamma/\Gamma_L \Gamma_R)^2 \TT^2$  
\cite{SuppInfo}.
We also find $\Delta_\phi = 1 + \RR\Delta_\theta$,
independent of the tunneling rates
and the detuning 
of the incident electron energy $\e$
from $\ed$.
Comparing with \eq{eq:damping}, 
this implies that the phases play a crucial role:  
{\it the $x$-dependence of the
overall scattering phase $\phi$
always accounts for half of the damping},
as seen in \fig{fig:rlm}b.
Further, in the limit of asymmetric
tunneling rates 
we find $\RR\Delta_\theta \gg 1$,
and the damping is 
almost entirely due to the combined 
contributions from
$\phi$ and $\theta$.
Also striking is the cotunneling limit,
where the detuning is large
compared to the level broadening,
i.e.~$\abs{\e - \ed} \gg \hbar\Gamma$.
In this limit tunneling is suppressed,
$\TT \ll 1$, and 
the level charge only fluctuates
virtually; as a result, one might expect that
the system is equivalent to a 
single junction in the weak tunneling limit, 
and
that $\gamma$ should be given by the
tunnel Hamiltonian result, i.e.~the 
first term in \eq{eq:damping}.
However, this is not the case:
due to phase corrections,
the damping is {\it twice} the
tunnel Hamiltonian result
(see inset of \fig{fig:rlm}b).
This shows that the phases can 
play a role even when $\TT$ is small.
Note that backaction in 
this model was recently studied
theoretically using a path integral
approach \cite{MozyrskyDisplace2004,Hussein10},
although backaction due to phases was
not discussed.

Our general theory also allows us to 
consider variations of the above RLM
where the mechanical position modulates 
the tunneling rates $\Gamma_L$ and $\Gamma_R$.  
This is the electromechanical analog
of a dissipatively coupled optomechanical
system \cite{Elste09},
and could be achieved 
experimentally
using a quantum dot coupled to two leads
via tunnel junctions, 
with the tunneling rates
modified 
by an on-board \cite{Flowers-Jacobs07}
or off-board \cite{Poggio08} mechanical oscillator.
First, we consider a setup
where only the left tunneling rate is $x$-dependent
(see \fig{fig:rlm}c,d).
In this case,
interference between resonant charge fluctuations
(on the level) and non-resonant
charge fluctuations (in the leads), result
in a Fano lineshape and suppression of $\gamma$ at
zero detuning \cite{SuppInfo},
similar to the optomechanical 
case \cite{Elste09}.
Second, we consider mechanical
coupling to both tunneling rates with
opposite sign, corresponding to
a quantum shuttle (see \fig{fig:rlm}e,f).
Here we find $\gamma \propto (\Gamma/\Gamma_L \Gamma_R)^2 \TT$;
moreover, all of the damping is due
to the scattering phases,
since (for $\Gamma_L = \Gamma_R$)
the transmission has no
linear dependence
on $x$ \cite{SuppInfo}.

{\it Atomistic model.}---While 
the above examples 
show that phases contribute to
backaction in simple 
model potentials, our approach allows us to 
investigate phase contributions in fully
atomistic calculations of mesoscopic conductors.
We demonstrate this by applying our theory to 
an APC
using the scattering matrix obtained from 
DFT  \cite{matdcal1}.
\begin{figure}[tb]
\centering
	\hspace{4mm}
	\includegraphics[width=0.3\textwidth]{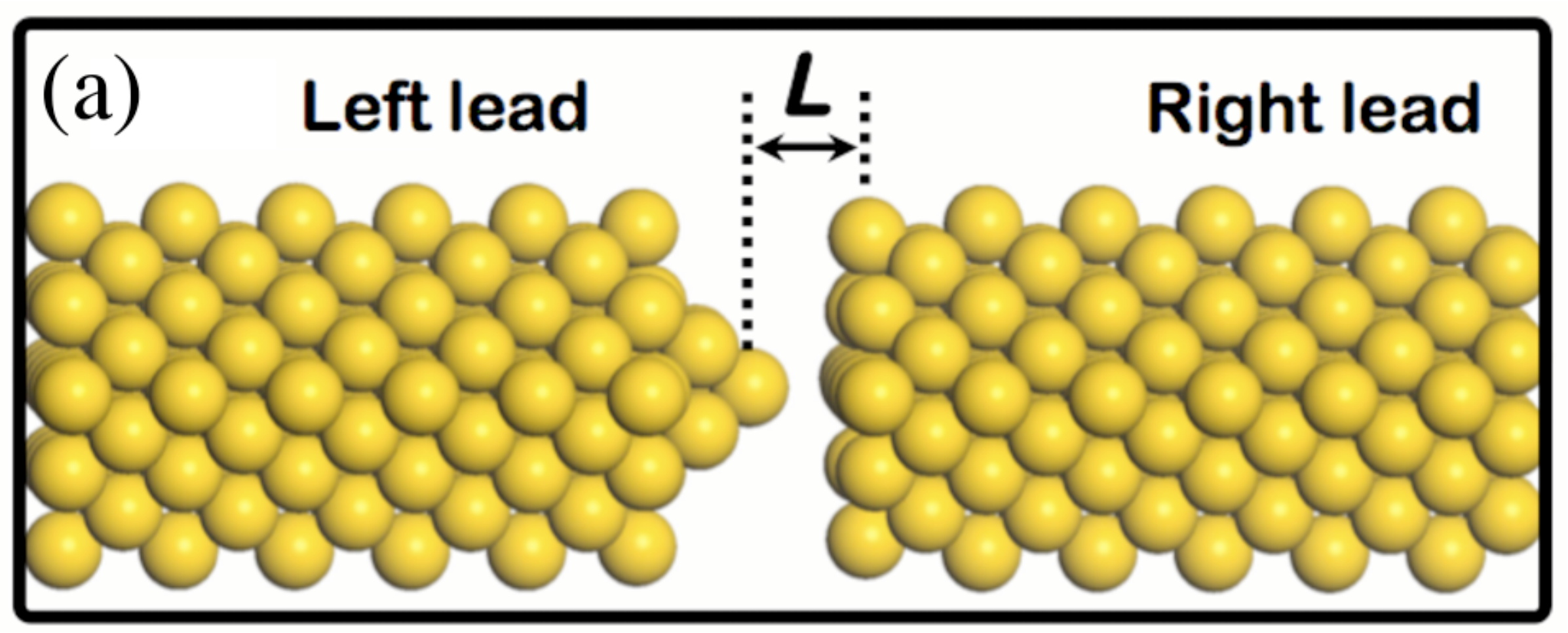}
	\includegraphics[width=0.4\textwidth]{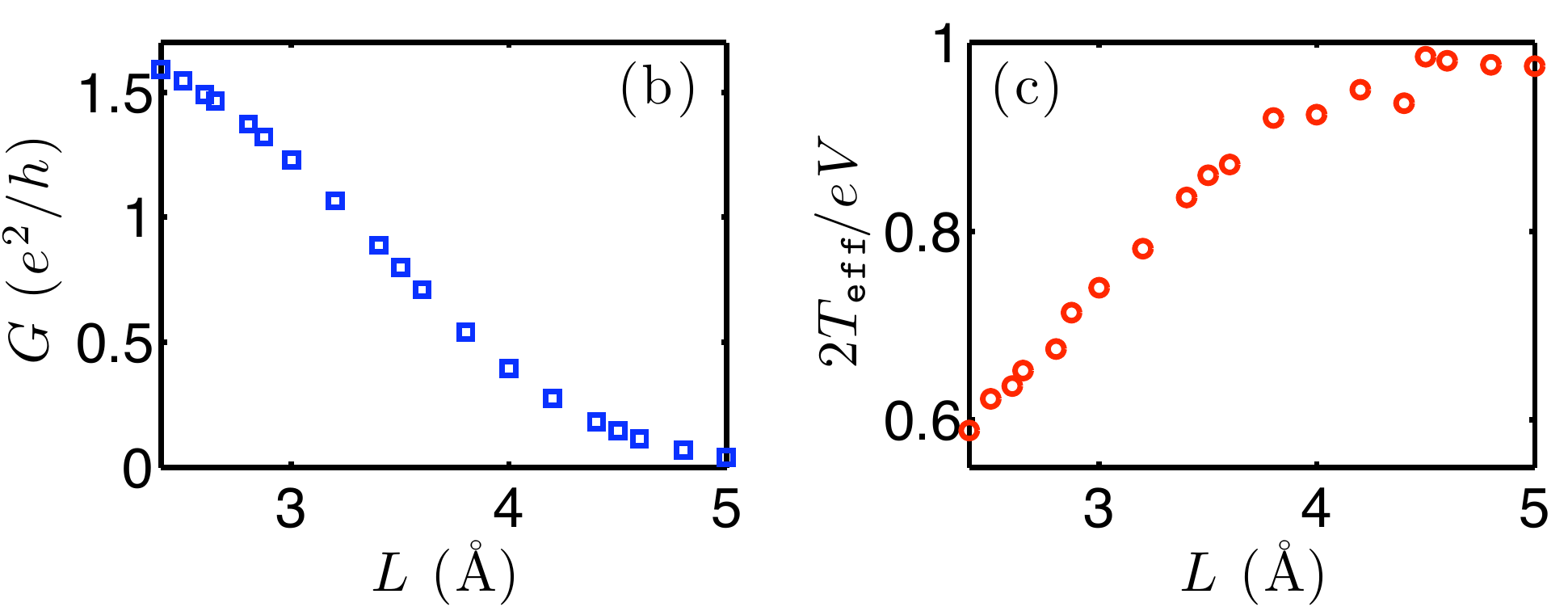}
\caption{(a) APC in a $5 \times 5$ atom Au (100) quantum wire.
(b) Conductance 
through the APC versus gap size
(for small $L$, several
channels contribute to 
transport and $G$ exceeds $e^2/h$ \cite{SuppInfo}).
(c) Effective backaction temperature
versus gap size.
For small $L$, $\Teff < eV/2$
due to phase corrections.
The Au-Au interatomic distance in
bulk gold is 2.87 \AA.
}
\label{fig:dft}
\end{figure}
We model the APC as a 
single-atom constriction
in a 5$\times$5 atom Au quantum wire
(see \fig{fig:dft}),
and take $x$ to modify the
gap size of the APC to $L+x$.
This geometry is
motivated by recent experimental 
setups using an
APC \cite{Flowers-Jacobs07}
or scanning tunneling microscope (STM) \cite{Kemiktarak07}
with one mechanically compliant electrode.
We approximate the surface electrode
of experiments by the flat
5$\times$5 edge of wire on the right;
this is justified since the 
transport properties
of the APC are 
expected to be
dominated by the few atoms
closest
to the tip.
We find 11 scattering channels
contributing to transport,
consistent with recent {\it ab initio} studies
of similar Au wires \cite{leads}.
After obtaining $s$ and $\partial s / \partial x$
\cite{SuppInfo},
we calculate the backaction using
\twoEq{eq:SFgen}{eq:dampingGen},
assuming $\Te \ll eV$.
We find that phase corrections are
important when the APC transmission 
deviates from the weak tunneling limit; 
it leads to a 
significant reduction in $\Teff$ from the 
tunnel Hamiltonian 
result of $eV/2$, as seen in 
\fig{fig:dft}.
While transmission properties 
are often studied using DFT, 
an important feature of our calculation
is our
explicit use of the scattering
phases obtained from an atomistic 
calculation of a quantum electronic
device.

{\it Conclusions.}---We have presented 
a scattering approach
to backaction in NEMS and
demonstrated the importance of backaction
from scattering
phases.
This work is particularly relevant to NEMS based
on quantum or atomic point contacts which
are often far from the weak tunneling limit.
Our results may also be 
easily extended to describe
strong electromechanical coupling in the
low oscillator frequency limit, by making an adiabatic
approximation such that the
noise spectra of $\hat F$ effectively
become $x$-dependent \cite{ClerkBennett05}.

This work was supported by
NSERC, FQRNT and CIFAR.


\clearpage
\renewcommand{\theequation}{E\arabic{equation}}
\setcounter{equation}{0}

\section{Erratum added}

In Ref.~\cite{BennettScattering10}, 
Eq.~(4) for the backaction damping
is valid only when
the scattering matrix $s(\e,x)$ is
independent of energy over the
range of energies contributing to scattering; 
we erroneously stated that this expression 
was correct for general, energy-dependent scattering.
Using the definition of the $W$ matrix given in Eq.~(1) 
of Ref.~\cite{BennettScattering10},  
one can easily show that the full expression 
for the low-frequency, linear-response backaction damping 
(valid for arbitrary energy-dependent scattering) 
is
\begin{widetext}
\begin{equation}
\label{eq:dampingFull}
	\gamma
	= 
	\frac{\pi\hbar}{M} \int d\e\
	\text{tr}
	\left\{
	\sum_{\alpha\beta}
	W_{\alpha\beta}W_{\beta\alpha}
	 \left( - \mypartial{f_\alpha}{\e} \right)
	 + 2 \left( f_L - f_R \right)
	 \left[
	 \mypartial{}{\omega}	
	 W_{LR}\left(\e-\frac{\omega}{2},\e+\frac{\omega}{2} \right) 
	 W_{RL}\left(\e+\frac{\omega}{2},\e-\frac{\omega}{2}\right)
	 \right]_{\omega=0}
	 \right\}.
\end{equation}	
\end{widetext}
We stress that 
this correction
has no impact on the subsequent
results of Ref.~\cite{BennettScattering10}, 
as
we exclusively considered
the limit of small drain-source voltages
in which the
energy dependence of
the scattering matrix
plays no role.

It is worth briefly outlining the different
origins of the terms in \eq{eq:dampingFull}.
The first term, 
proportional to $W_{\alpha\beta}W_{\beta\alpha}$
evaluated at energy $\e$,
corresponds to the damping
discussed in Ref.~\cite{BennettScattering10}.
This term arises from the 
increase in the number of
scattering transitions 
contributing to the force noise when 
an electron 
in the conductor absorbs
energy $\omega$ from the mechanical
degree of freedom
(and conversely, the decrease if an electron emits
energy to the mechanics).
Because of this asymmetry, the conductor favors
absorption of energy and this part of the 
damping is always positive.
In contrast, the term in \eq{eq:dampingFull}
proportional to $f_L - f_R$
constitutes a nonequilibrium
contribution originating from the energy dependence
of the matrix elements of the force operator 
between scattering states at different energies.
This intrinsic energy dependence 
of the matrix elements
may favor either absorption or emission
of energy resulting in positive or negative damping.
This contribution was recently presented in 
Ref.~\cite{Bode11} for a general model where the 
electronic system is a multilevel, noninteracting  quantum dot.

It is interesting to note that, following the 
lines of Ref.~\cite{Smith60}, it is not possible to relate the 
nonequilibrium term in $\gamma$ to the ``frozen" 
scattering matrix $s(\e,x)$; 
one also needs knowledge 
of the form of the wavefunctions in the scattering region. 
Nonetheless,
one can easily identify general
classes of systems where it plays no 
role.
We find that the nonequilibrium term always 
vanishes for an 
inversion-symmetric scattering potential, 
as well as for the single-resonant-level 
models considered in Ref.~\cite{BennettScattering10}.  The non-equilibrium 
damping terms thus play no role 
(even at finite bias voltage) for the 
examples considered in Ref.~\cite{BennettScattering10}.

We 
thank F.~von Oppen for drawing our 
attention to the incorrect presentation 
of Eq.~(4) in our original paper.


\end{document}


\title{Scattering approach to
backaction in coherent nanoelectromechanical systems:
\\
Supplemental Information}
\author{Steven D. Bennett}
\author{Jesse Maassen}
\author{Aashish A. Clerk}
\date{\today}
\affiliation{Department of Physics, McGill University,
Montreal, Quebec, Canada, H3A 2T8}

\maketitle

\section{Resonant level model}

As given in the main text,
the scattering matrix for the RLM 
(see, for example, \cite{ButtikerChristen96}) is
\begin{equation}
	s_{\alpha\beta}(\e,x)
	= \delta_{\alpha\beta} - 
	\frac{ i \sqrt{ \Gamma_\alpha \Gamma_\beta }}
	{ \e - \ed + i \Gtot/2 },
\end{equation}
where 
$\ed$ is the level energy,
$\Gamma_L$ ($\Gamma_R$) is
the tunneling rate to the left (right)
lead and
$\Gtot = \Gamma_L + \Gamma_R$.
Here we set $\hbar = 1$.
The transmission probability
is $\TT = \Gamma_L \Gamma_R / \left[ (\e - \ed))^2 + \Gtot^2/4 \right]$.
In Fig.~1 of the main text we 
sketch and plot the damping for
three forms of mechanical coupling.
First, 
dispersive 
coupling corresponds to
an $x$-dependent 
level energy and we replace 
$\ed$ with $\tilde \e_d(x)$.
We assume linear coupling,
$\tilde \e_d(x) =  \ed + A x$, and
obtain
\begin{equation}
	\gamma = \frac{A^2}{4\pi M}
	\frac{\Gtot^2}{ \left[  (\e - \ed)^2 + \Gtot^2/4 \right]^2}.
\end{equation}
and
we see that 
$\gamma \propto (\Gamma / \Gamma_L \Gamma_R)^2 \TT^2$
as stated in the main text.
Second, dissipative coupling corresponds
to $x$-dependent tunneling rates.
We take the left rate to depend linearly
on $x$ and
replace $\Gamma_L$ with
$\tilde\Gamma_L(x) = \Gamma_L + A x$.
This yields
\begin{equation}
\label{eq:dampingDiss}
	\gamma = \frac{A^2}{4 \pi M}
	\frac{\left( \Gtot + \Gamma_L \right) (\e-\ed)^2 + \Gamma_R \Gtot^2/4}
	{2\Gamma_L \left[ (\e-\ed)^2 + \Gtot^2/4 \right]^2}.
\end{equation}
This result can be verified
by a direct calculation starting from
the Hamiltonian for a single resonant level,
and using the same $\tilde\Gamma_L(x)$
\cite{BennettPhD}.
Finally, we model a quantum shuttle with
$x$-dependent left and right tunneling rates,
$\tilde\Gamma_L(x) = \Gamma_L + A x$ and
$\tilde\Gamma_R(x) = \Gamma_R - A x$.
We find
\begin{equation}
\label{eq:dampingShutt}
	\gamma = 
	\frac{A^2}{4\pi M}
	\frac{\Gamma^2}
	{\Gamma_L \Gamma_R \left[ (\e-\ed)^2 + \Gtot^2/4 \right]},
\end{equation}
and we see that 
$\gamma \propto (\Gamma / \Gamma_L \Gamma_R)^2 \TT$
as stated in the main text.

\section{Atomic point contact model}

\subsection{Atomic structure and {\em ab initio} methods}

The APC consists of a tapered 5$\times$5 
atom Au(100) quantum wire opposing an identical 
flat-surfaced nanowire (see Fig.~2 in the main text).
The atomic positions of the three Au layers forming the 
tip (on the left) and the first Au layer of 
the flat surface (on the right) were 
independently relaxed 
to forces 
$< 0.01\ \text{eV/\AA}$ 
using the VASP
density functional theory (DFT) software package \cite{vasp1,vasp2}.
All non-relaxed atoms were fixed to their DFT optimized
bulk positions corresponding to a lattice constant of
4.06 \AA\ for FCC gold. 

From the optimized APC structure, we performed 
the first principles transport calculations 
using the {\sc MatDCal} device simulator, based on a
combination of DFT and Keldysh nonequilibrium Green's 
functions (NEGF) \cite{matdcal1,matdcal2}. {\sc MatDCal} uses a
linear combination of atomic orbitals as a basis, where
two s-orbitals, one p-orbital and two d-orbitals 
were employed for each Au atom. The exchange and 
correlation energies were described within 
the local density approximation \cite{lda},
while the nuclear and core electrostatic
potentials were modelled using norm-conserving
non-local pseudopotentials \cite{pseudo}.
Note that the size of the supercell box,
in the directions perpendicular to the transport
direction, was chosen large enough 
to ensure no interactions with the 
neighboring supercells. After calculating the self-consistent
DFT Hamiltonian, we solve for the scattering
matrix which provides the transmission and reflection
amplitudes.


\subsection{Transmission eigenvalues}

\begin{figure}[tb]
\centering
	\includegraphics[width=0.4\textwidth]{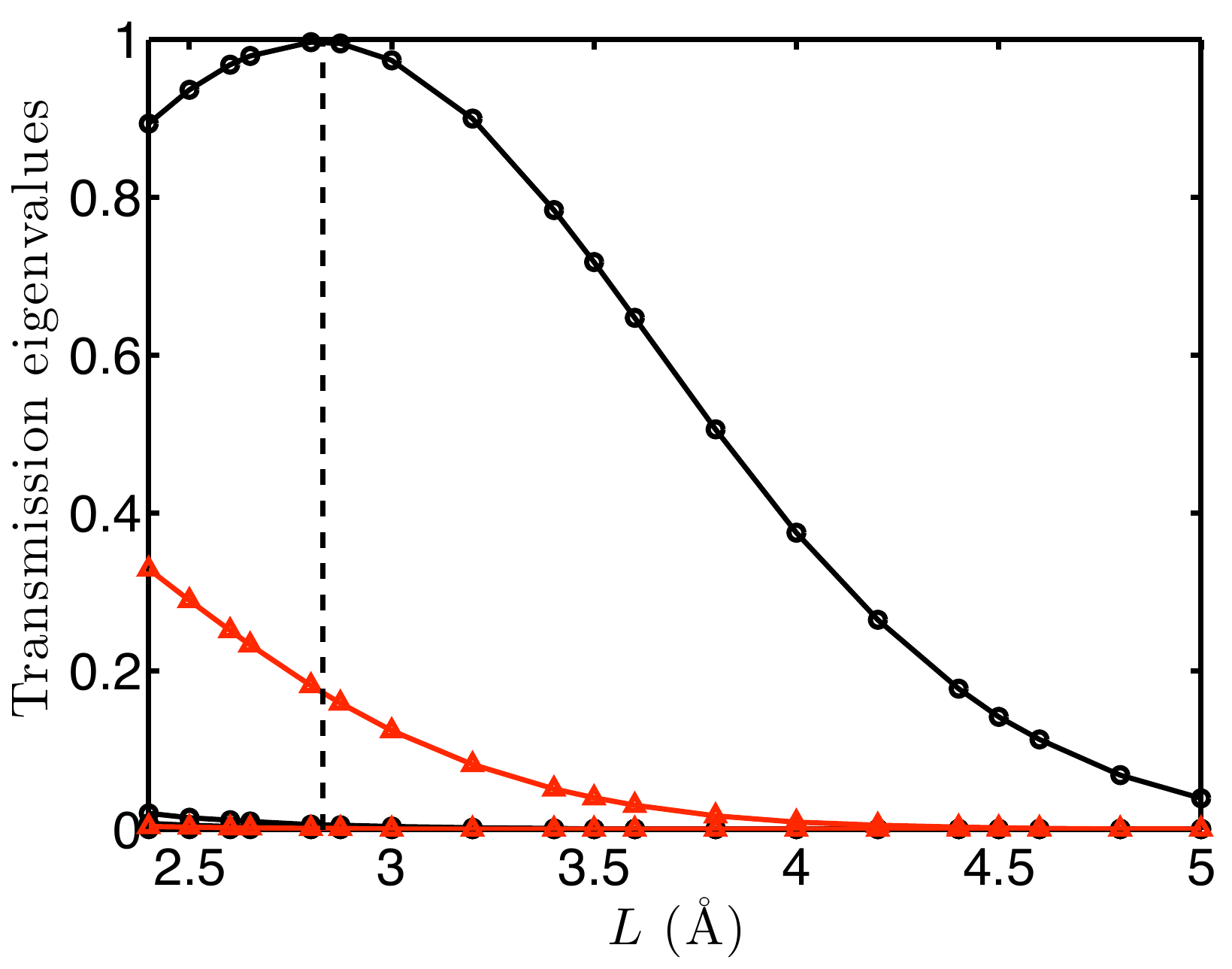}
\caption{Transmission eigenvalues versus
APC gap size.
Red triangles
indicates two-fold degenerate eigenvalues.
The vertical dashed line marks 
the Au-Au 
interatomic distance
of 2.87 \AA\
in bulk gold.
}
\label{fig:Teig}
\end{figure}

Our calculation yields 11 scattering channels
(i.e. 11 propagating states at the Fermi level) 
in the 5$\times$5 atom
Au(100) quantum wire, consistent
with previous calculations \cite{leads}.
We obtain the  transmission eigenvalues 
by diagonalizing the $11 \times 11$ transmission
block of the scattering matrix connecting the
left and right leads, $s_{LR}$.
The eigenvalues
are shown in \fig{fig:Teig} versus
the APC gap size.
One transmission eigenvalue
dominates the others at 
all gap sizes, and becomes
perfectly conducting
when the atomic gap size
is near the Au-Au interatomic distance of 2.87 \AA.
We also find two sets of two-fold degenerate 
transmission eigenstates
(red triangles) for all gap sizes.

\subsection{Damping and force noise}

\begin{figure}[tb]
\centering
	\includegraphics[width=0.45\textwidth]{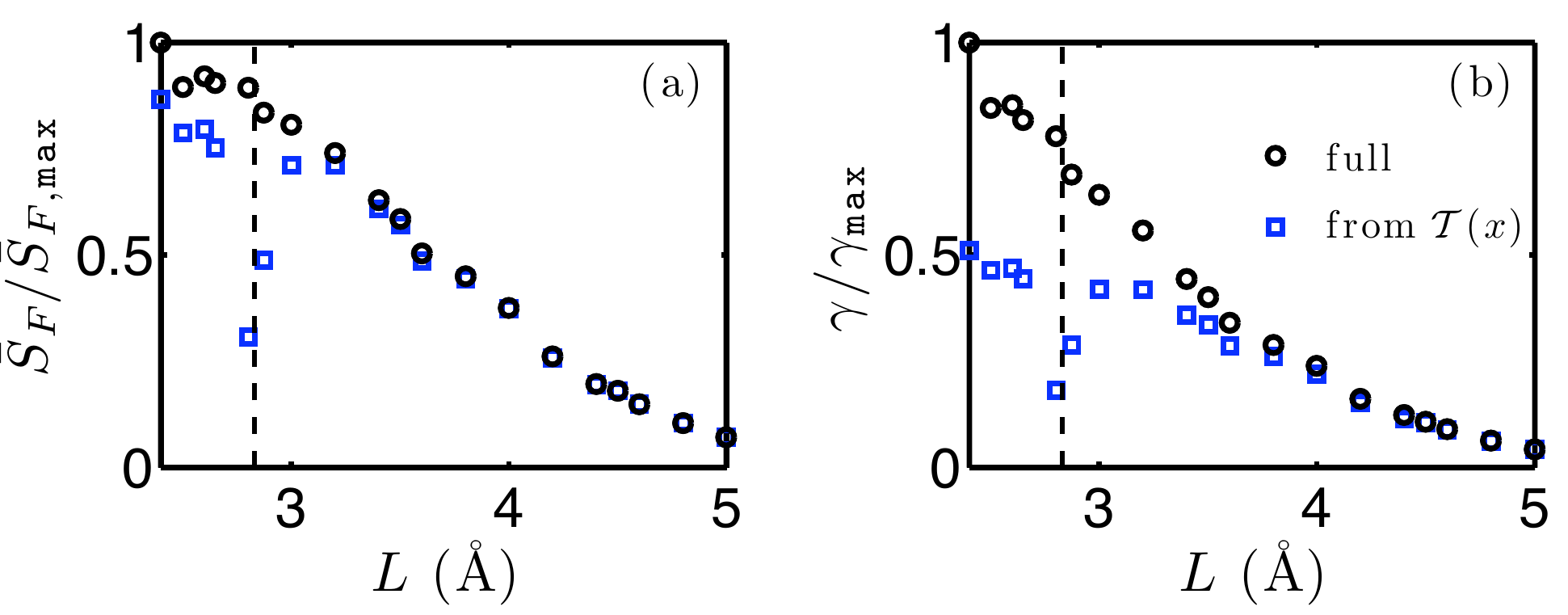}
\caption{Damping and force noise
versus APC gap size,
calculated from the scattering
matrix obtained using DFT.
Dips appear at the Au-Au 
interatomic distance
of 2.87 \AA\
in bulk gold
(vertical dashed line)
in both quantities calculated from
the $x$-dependence of transmission only;
however, no dips are present when
phase corrections
are included.
Both $\gamma$ and $\bar S_F$ 
are scaled by their maximum
values (from the full calculation).
}
\label{fig:gammaSF}
\end{figure}

To calculate the effective temperature plotted in
Fig. 2c of the main text, we first calculate
the symmetrized force noise and
damping from the scattering
matrix
using Eqs.~(3) and (4) of the main text.
$\Teff$ is then obtained from 
$\Teff = \bar S_F / 2M \gamma$
as discussed in the main text.
The force noise and damping
are shown in \fig{fig:gammaSF}.
For comparison, we also show the 
force noise and damping calculated
from the $x$-dependence of the
transmission only, neglecting
the phase contributions.
Both quantities calculated
from $\TT(x)$ alone show
dips at a gap size 
near the Au-Au interatomic distance
in bulk gold.
This is not surprising,
since this point corresponds
to a minimum in the
total free energy
of the atomic system; moreover,
the dominant
transmission eigenvalue
reaches a maximum at this point
(see \fig{fig:Teig}).
Interestingly,
corrections from the
scattering phases completely
wash out these dips.
Note that the dips are not reflected
in $\Teff$;
from the transmission terms only,
we always obtain $\Teff = eV/2$.
